\documentclass[11 pt]{article}
\usepackage{authblk}
\usepackage[]{inputenc}
\usepackage{amsfonts}
\usepackage{enumitem}
\usepackage{latexsym,amsmath}
\usepackage{bigints}
\usepackage{amssymb,array}
\usepackage{xcolor}
\usepackage{hyperref}
\usepackage{bm}
\usepackage{multirow}
\hypersetup{
	colorlinks=true,
	linkcolor=blue,
	citecolor=blue,
	filecolor=magenta,      
	urlcolor=red,
}
\usepackage{caption}
\usepackage{subcaption}
\usepackage{graphics}
\usepackage{epstopdf}
\usepackage{array}
\makeatletter \oddsidemargin 0in \evensidemargin 0in \textwidth 16cm
\RequirePackage[dvips]{graphicx} \textheight 20cm
\begin{document}
	\newcommand{\wt}{\widetilde}
	\newcommand{\lt}{\left}
	\newcommand{\rt}{\right}
	\newcommand{\ap}{\alpha}
	\newcommand{\bt}{\beta}
	\newtheorem{thm}{Theorem}[section]
	\newtheorem{pro}{Proposition}[section]
	\newtheorem{r1}{Remark}[section]
	\newtheorem{ce}{Counterexample}[section]
	\newtheorem{cor}{Corollary}[section]
	\newtheorem{d1}{Definition}[section]
	\newtheorem{ex}{Example}[section]
	\newtheorem{lem}{Lemma}[section]
	\numberwithin{equation}{section}
	\title{On multivariate orderings of some general \\ordered random vectors}
	\author[a]{Tanmay Sahoo}
	\author[a]{Nil Kamal Hazra  \footnote{Corresponding author. Email: nilkamal@iitj.ac.in}}
	\author[b]{Narayanaswamy Balakrishnan}
	\affil[a]{Department of Mathematics, Indian Institute of Technology Jodhpur, Karwar 342037, India}
	\affil[b]{Department of Mathematics and Statistics, McMaster University, Hamilton, Ontario, Canada L8S 4K1}
	\date{}
	\maketitle
	\begin{abstract}
	Ordered random vectors are frequently encountered in many problems. The generalized order statistics (GOS) and sequential order statistics (SOS) are two general models for ordered random vectors. However, these two models do not capture the dependency structures that are present in the underlying random variables. In this paper, we study the developed sequential order statistics (DSOS) and developed generalized order statistics (DGOS) models that describe the dependency structures of ordered random vectors. We then study various univariate and multivariate ordering properties of DSOS and DGOS models under Archimedean copula. We consider both one-sample and two-sample scenarios and develop corresponding results.
	\end{abstract}
	\\\hspace*{3.2 in}Secondary 60E15; 60K10
	\section{Introduction}\label{se110}
	Order statistics (OS) and record values arise naturally in several statistical modeling and
inferential problems (see \cite{ABN, ABN1, HC, J, KC}). As a more general framework in which both these models are incorporated, the
notion of generalized order statistics (GOS) was introduced. {In addition}, this GOS model contains
several other models of ordered random variables, {such as}, order
statistics with non-integral sample size, $k$-record values, Pfeifer’s records, $k_n$-records
from non-identical distributions, ordered random variables from truncated 
distributions, progressively type-II censored order statistics, and so on. 
Thus, the GOS model
provides a unified class of models, with a variety of {interesting} and practical characteristics, which can be used to describe and study many real-world problems. On the other hand, the sequential
order statistics (SOS), an extension of ordinary order statistics (OS), are used to represent the
lifetimes of systems. In the SOS model, the failure of any component has an impact on the remaining surviving components and so the distributions of the lifetimes of
remaining components are assumed to differ from the original ones. In reliability theory, there is a one-to-one relation between SOS and the lifetimes of sequential $k$-out-of-$n$ systems (see the definition in \cite{CK}). In fact, the lifetime of a sequential $k$-out-of-$n$ system is the same as the $(n - k + 1)$-th sequential order statistic of the lifetimes of components of the system. One may note that the GOS model is closely related to the SOS model. In particular, a specific choice of distribution functions (i.e., under the proportional hazard rate (PHR) model) in the SOS model leads to the GOS model. Thus, the SOS model can be viewed as a more generalized model that contains almost all existing models of ordered random variables.
\\\hspace*{0.2 in}In the literature, numerous studies have been carried out concerning univariate and multivariate stochastic comparisons of ordinary order statistics (see \cite{ BZ, BP, BRR, BGR, HKFN, LF, NLP, NM,  NR, SH, SAE} and the references therein). In the same vein, stochastic comparisons of generalized order 
statistics as well as stochastic comparisons of sequential order statistics have been discussed in the literature. Belzunce et al.~\cite{BMR} developed several results concerning multivariate and univariate stochastic comparisons of generalized order statistics with respect to the usual stochastic order, dispersive order, hazard rate order and likelihood ratio order. Hu and Zhuang~\cite{HZ} subsequently added some more results on univariate stochastic comparisons of generalized order statistics. Chen and Hu~\cite{CH} studied ordering properties of generalized order statistics with respect to the multivariate dispersive order. 
Xie and Hu~\cite{XH} subsequently discussed stochastic comparisons of multivariate marginals of generalized order statistics with respect to multivariate dispersive order. Balakrishnan et al.~\cite{BBSS} derived some results for stochastic comparisons of generalized order statistics with respect to increasing convex order. Some more works on generalized order statistics can be found in \cite{TA, BM}, and the references therein. Additionally, the study of various univariate orderings and ageing properties of sequential order statistics has been carried out by \cite{BN2, BT, NB, TLW}. Zhuang and Hu~\cite{ZH} studied multivariate stochastic comparisons of sequential order statistics with respect to multivariate likelihood ratio order, multivariate hazard rate order and multivariate usual stochastic order. One may note that all the studies listed above, for GOS and SOS models, have been carried out under the assumption that the underlying random variables are independent. 
\\\hspace*{0.2 in}The SOS model is defined based on
the assumption that the lifetimes of the set of remaining components in each step (i.e., after each failure)
are independent. This is indeed a very stringent assumption in many real-life scenarios. For example, consider the oil transmission pipeline station with five pumps. Suppose the station functions effectively as long as three out of the five pumps are operational. Here, the lifetimes of the five pumps are indeed dependent, and the failure of a pump increases the load on the remaining pumps because a proper transmission requires a certain level of oil pressure (i.e., load-sharing effect). This is an example of a sequential $3$-out-of-$5$ system with dependent component lifetimes (see \cite{BD}). 
\\\hspace*{0.2 in}To overcome the aforementioned drawback of the SOS model, Baratnia and Doostparast~\cite{BD} recently introduced the notion of developed sequential order statistics (DSOS), which is an extended SOS model. The DSOS model captures the dependency structure between components of a
system in each step. 
Recently, Sahoo and Hazra~\cite{SH1} have studied various univariate stochastic comparison results for DSOS wherein the dependency structure has been described by an Archimedean copula. 
However, no study has been carried out for multivariate stochastic comparisons of DSOS.
Thus, one of our main goals in this paper is to study various univariate and multivariate stochastic comparisons of DSOS governed by an Archimedean copula. {In analogy} to DSOS model, we introduce the notion of developed generalized order statistics (DGOS), which is a GOS model involving dependent random variables. 
In particular, what  we study in this paper are the following:
\begin{itemize}
\item Various multivariate stochastic orderings (namely, multivariate
usual stochastic order, dynamic multivariate hazard rate order, and multivariate dispersive order)
and univariate stochastic orderings (namely, usual stochastic order, hazard rate order, reverse
hazard rate order, dispersive order, and increasing convex order) properties of developed
sequential order statistics (DSOS) and developed generalized order statistics (DGOS) in
both one and two-sample situations.
\end{itemize}
It is worthwhile to
mention that the results established here generalize many known results on sequential order statistics, generalized order statistics, record values,  progressively type-II censored order statistics, order statistics from truncated distributions, and usual order statistics. The
novelty {in this work} is mainly in considering the DGOS and DSOS models based on Archimedean copula. 
\\\hspace*{0.2 in}The rest of {this} paper is organized as follows. In Section~\ref{se111}, we {present} some preliminaries. In Section~\ref{se111a}, we discuss the notion of some ordered random vectors. In Section~\ref{se113}, we establish some stochastic comparison results for random vectors from DSOS model with identical components. In Section~\ref{se114}, we establish some stochastic comparison results for DGOS model with identical components. 
Finally, some concluding remarks are made in Section~\ref{se116}. 
\section{Preliminaries}\label{se111}	
\hspace*{0.2 in} Unless otherwise stated, we use the following notation throughout the paper. For an absolutely continuous random variable Z, we denote the cumulative distribution
function (CDF) by $F_Z(\cdot)$, the reliability function (RF) by $\bar{F}_Z(\cdot)$, the probability density function (PDF) by $f_Z(\cdot) $, {and the
cumulative hazard rate function by $\Delta_{Z}(\cdot)$, where $\bar{F}_Z(\cdot) \equiv 1-F_Z(\cdot)$ and $\Delta_{Z}(\cdot)  \equiv  -\ln \bar{F}_Z(\cdot)$.
We denote the set of natural numbers and the set of real numbers by $\mathcal{N}$ and $\mathcal{R}$, respectively. {We write} $a\stackrel{d}=b$ {to} mean that $a$ and $b$ have the same distribution.
\\\hspace*{0.2 in}
Copulas are very useful in describing the dependence structure between random variables.
A wide range of copulas have been discussed in the literature {and some} of the well-known copulas {are} Farlie-Gumbel-Morgenstern (FGM) copula, extreme-value copula, Archimedean copulas, and Clayton-Oakes (CO) copula. The family of Archimedean copulas have received considerable attention {due} to their {tractability} and ability to capture a wide range of {dependence.} A comprehensive description of this topic can be found in the book by Nelsen~\cite{N}. Below, we give the definition of {an} Archimedean copula (see~\cite{MN}).
\begin{d1}
Let $\phi: [0, +\infty]\longrightarrow [0,1]$ be a decreasing continuous function with $\phi(0)=1$ and $\phi(+\infty)=0$, {and} $\psi\equiv\phi^{-1}$ be the pseudo-inverse of $\phi$. Then,
	\begin{eqnarray}
	C(u_1,\dots,u_n)=\phi\lt(\psi(u_1)+\dots +\psi(u_n)\rt),\quad \text{for }(u_1,\dots,u_n)\in [0,1]^n,
	\end{eqnarray}
	is called an Archimedean copula with generator $\phi$ if $(-1)^k\phi^{(k)}(x)\geq 0$, for $k=0,1,\dots,n-2$, and  $(-1)^{n-2}\phi^{(n-2)}(x)$ is decreasing and convex in $x\geq 0$, where $\phi^{(k)}(\cdot)$ represents the $k$-th derivative of $\phi$.$\hfill\Box$
\end{d1}
We now introduce some key notation that will be used in the sequel. For an Archimedean copula with generator $\phi$, we denote
	\begin{eqnarray*}
&& H(u)=\frac{u\phi'(u)}{1-\phi(u)},\;\;R(u)=\frac{u\phi'(u)}{\phi (u)}\text{ and }G(u)=\frac{u\phi''(u)}{\phi'(u)},\quad u>0.
\end{eqnarray*}
Note that $H(\cdot)$, $R(\cdot)$ and $G(\cdot)$ are all negative-valued functions since $\phi(\cdot)$ is a decreasing convex function.
\\\hspace*{0.2 in}Before proceeding further, we introduce the following notation. For cumulative distribution functions $F_i$, $i =1,2,\dots,n$, we denote the {corresponding} probability density functions, quantile functions, survival functions, hazard rate functions, reversed hazard rate functions, and cumulative hazard rate functions by  $f_i$, $F_i^{-1}$, $\bar F_i$, $r_i$, $\tilde r_i$ and $D_i$, respectively, where $r_i\equiv f_i/\bar F_i$, $\tilde r_i\equiv f_i/F_i$ and $D_i\lt(\cdot\rt)\equiv -\ln\bar{F}_i\lt(\cdot\rt)$. Similarly, for cumulative distribution functions $G_i$, $i =1,\dots,n$, we denote the {corresponding} probability density functions, quantile functions, survival functions, hazard rate functions, reversed hazard rate functions and cumulative hazard rate functions by  $g_i$, $G_i^{-1}$, $\bar G_i$, $h_i$, $\tilde h_i$ and $B_i$, respectively, where $h_i\equiv g_i/\bar G_i$, $\tilde h_i\equiv g_i/G_i$ and $B_i\lt(\cdot\rt)\equiv -\ln\bar{G}_i\lt(\cdot\rt)$.
\\\hspace*{0.2 in}The proportional hazard rate (PHR) model is one of the commonly used semi-parametric models in survival analysis and reliability theory. A set of random variables $\{Z_1,\dots,Z_n\}$ is said to follow the PHR model if, for $i=1,\dots,n$, 
$$\bar F_{Z_i}(t)=(\bar F(t))^{\alpha_i}, \text{ for some }\alpha_i>0\;\;\text{and for all }t>0,$$	
where $\bar F$ is the baseline survival function. We {shall} denote this by 
$F_{Z_i}\sim$ PHR($ F; \alpha_i$), for $i=1,\dots,n$. 
\\\hspace*{0.2 in}Stochastic orders are very effective tools for comparing two or more random variables/vectors. 
Below, we give the definitions of {some} stochastic orders (see~\cite{SS}) that are most pertinent to the subsequent discussion. 
 \begin{d1}
	Let $X$ and $Y$ be two absolutely continuous random variables with non-negative supports. Then, $X$ is said to be smaller than $Y$ in the
		\begin{enumerate}[noitemsep]
		\item [$(a)$] usual stochastic order, denoted by $X\leq_{st}Y$ or $F_X \leq_{st} F_Y$, if $\bar F_X(x)\leq \bar F_Y(x) \text{ for all }x $   $\in~[0,\infty);$ 
		\item [$(b)$] hazard rate order, denoted by $X\leq_{hr}Y$ or $F_X \leq_{hr} F_Y$, if ${\bar F_Y(x)}/{\bar F_X(x)}\text{ is increasing in } x \in [0,\infty);$
		\item [$(c)$] reversed hazard rate order, denoted by $X\leq_{rh}Y$ or $F_X \leq_{rh} F_Y$, if $ {F_Y(x)}/{ F_X(x)}\text{ is increasing} $  $\text{ in } x\in [0,\infty);$
		\item [$(d)$] likelihood ratio order, denoted by $X\leq_{lr}Y$ or $F_X \leq_{lr} F_Y$, if
		${f_Y(x)}/{f_X(x)}\;\text{ is increasing } $  $\text{ in } x\in(0,\infty);$
		\item [$(e)$] dispersive order, denoted by $X\leq_{ disp} Y$ or $F_X \leq_{disp} F_Y$, if
		$G^{-1}(u) - F^{-1}(u)$ is increasing in $u \in (0,1)$;
		\item [$(f)$] increasing convex order, denoted by $X\leq_{icx}Y$ or $F_X \leq_{icx} F_Y$, if $E(\phi(X)) \leq E(\phi(Y))$, for all increasing convex functions $\phi$;
		\item [$(g)$] mean residual life order, denoted by $X\leq_{ mrl} Y$ or $F_X \leq_{mrl} F_Y$, if
		$\int_{x}^{\infty}\bar{F}_Y(u)du/\int_{x}^{\infty}\bar{F}_X(u)du$ is increasing in $x$ over $\{x:\int_{x}^{\infty}\bar{F}_X(u)du >0\};$
		\item [$(h)$] ageing faster order in terms of hazard rate, denoted by $X \leq_{c} Y$ or $F_X \leq_{c} F_Y$, if {$ \Delta_X \circ \Delta_Y^{-1} $} is convex on $[0,\infty)$, or equivalently, $r_X /r_Y$ is increasing on $ [0, \infty)$. $\hfill\Box$
	\end{enumerate}
\end{d1}
 We now introduce the following notation. Let $\mbox{\boldmath$X$} = (X_1, \dots ,X_m)$ be a nonnegative random vector with an absolutely continuous distribution function. Consider a typical history of $\mbox{\boldmath$X$}$ at time $t \geq 0$, which is of the form
$$h_t = \{ \mbox{\boldmath$X$}_I = \mbox{\boldmath$t$}_I,\mbox{\boldmath$X$}_{\bar I} > t\mbox{\boldmath$e$}\}, 0\mbox{\boldmath$e$} \leq \mbox{\boldmath$t$}_I \leq t\mbox{\boldmath$e$}, I \subset \{1,\dots,m\};$$
here, $\mbox{\boldmath$t$}_I =(t_{i_1}, \dots, t_{i_k})$, $\bar I$ is the complement of $I= (i_1, \dots, i_k)$ in $\{1,\dots,m\}$ and $\mbox{\boldmath$e$}= (1,\dots,1)$. Given the history $h_t$, let $i \in \bar I$ be a component that is still alive at time $t$. Its multivariate conditional hazard rate, at time $t$, is defined as follows:
$$\lambda_{i | I}\lt(t | \mbox{\boldmath$t$}_I \rt)= \lim_{\Delta t \rightarrow 0^+} \frac{1}{\Delta t} P\lt( t < T_i \leq t+\Delta t | \mbox{\boldmath$T$}_I = \mbox{\boldmath$t$}_I, \mbox{\boldmath$T$}_{\bar I} >t\mbox{\boldmath$e$} \rt),$$
where, $0\mbox{\boldmath$e$} \leq \mbox{\boldmath$t$}_I \leq t\mbox{\boldmath$e$}$, and $I \subset \{1,\dots,m\}$ (see \cite{SS}).
\\\hspace*{0.2 in}Further, let $F_1$ be the marginal distribution function of $X_1$, {and} $F_{i |1,\dots,i-1}\lt(\cdot | x_1, \dots , x_{i-1}\rt)$ be the conditional distribution function of $X_i$, given $X_1 = x_1 , \dots, X_{i-1} = x_{i-1}$, for $i = 2,\dots,n$. For each $ \mbox{\boldmath$u$}= (u_1,\dots,u_n) \in (0,1)^n$, define
$$x_1\lt(\mbox{\boldmath$u$}\rt) = F_1^{-1}\lt(u_1\rt)$$
and {sequentially}
$$x_i \lt(\mbox{\boldmath$u$}\rt) = F^{-1}_{i | 1, \dots, i-1} \lt( u_i | x_1, \dots ,x_{i-1} \rt), \quad i=2,\dots,n.$$ 
\\\hspace*{0.2 in}Next, we present the definitions of some multivariate stochastic orders that are used in {the} subsequent sections.
 \begin{d1}
	 Let $\mbox{\boldmath$X$}$ and $\mbox{\boldmath$Y$}$ be two $n$-dimensional random vectors with non-negative supports. Further, let the multivariate probability density {functions} and the multivariate conditional hazard rate {functions} of $\mbox{\boldmath$X$}$ and $\mbox{\boldmath$Y$}$ be given by $f(\cdot)$ and $g(\cdot)$, and $\eta_{\cdot | \cdot}\lt( \cdot | \cdot\rt)$ and $\lambda_{\cdot | \cdot}\lt( \cdot | \cdot \rt)$, respectively. Then, $\mbox{\boldmath$X$}$ is said to be smaller than $\mbox{\boldmath$Y$}$ in the
		\begin{enumerate}[noitemsep]
		\item [$(a)$] usual multivariate stochastic order, denoted by $\mbox{\boldmath$X$} \leq_{st} \mbox{\boldmath$Y$}$, if $E\lt(\phi\lt( \mbox{\boldmath$X$} \rt)\rt) \leq E\lt(\phi\lt(\mbox{\boldmath$Y$}\rt)\rt)$, for all increasing functions $\phi$;
		\item [$(b)$] dynamic multivariate hazard rate order, denoted by $\mbox{\boldmath$X$} \leq_{dyn-hr} \mbox{\boldmath$Y$}$, if 
			\begin{eqnarray*}
	\eta_{k | I \cup J}\lt( u | \mbox{\boldmath$s$}_{I \cup J}\rt) &\geq&
	  \lambda_{k | I}\lt( u | \mbox{\boldmath$t$}_I\rt), \text{ for all } k \in \overline{I \cup J},
\end{eqnarray*}
where $I \cap J = \emptyset$, $\mbox{\boldmath$s$}_I \leq \mbox{\boldmath$t$}_I \leq u\mbox{\boldmath$e$}$ and $\mbox{\boldmath$s$}_J \leq u\mbox{\boldmath$e$}$; 
	\item [$(c)$] multivariate likelihood ratio order, denoted by $\mbox{\boldmath$X$} \leq_{lr} \mbox{\boldmath$Y$}$, if $f\lt( \mbox{\boldmath$x$} \rt) g\lt( \mbox{\boldmath$y$} \rt) \leq f\lt( \mbox{\boldmath$x$} \wedge \mbox{\boldmath$y$}  \rt) g\lt( \mbox{\boldmath$x$} \vee \mbox{\boldmath$y$}  \rt)$, for all $\mbox{\boldmath$x$}, \mbox{\boldmath$y$} \in \mathbb{R}^n$;
	\item [$(d)$] multivariate dispersive order, denoted by $\mbox{\boldmath$X$} \leq_{disp} \mbox{\boldmath$Y$}$, {if} $y_i\lt(\mbox{\boldmath$u$}\rt) - x_i\lt(\mbox{\boldmath$u$}\rt)$ is increasing in $(u_1,\dots, u_i) \in (0,1)^i$ for $i = 1,\dots,n$. $\hfill\Box$
		\end{enumerate}
\end{d1}
\hspace*{0.2 in}Like stochastic orders, majorization orders are also quite useful for establishing various  inequalities.  Different majorization orders have been {discussed} in the literature, and we give below the definitions of some majorization orders that are used in this work.
\begin{d1}
Let $I^n$ denote an $n$-dimensional Euclidean space, where $I\subseteq\mathcal{R}$. Further, let {$\mbox{\boldmath $x$}=(x_1,\dots,x_n)\in I^n$} and {$\mbox{\boldmath $y$}=(y_1,\dots,y_n)\in I^n$} be any two 
vectors, and {$x_{(1)} \le\cdots\le x_{(n)}$} and {$y_{(1)}\le\cdots\le y_{(n)}$} be the
increasing arrangements of the components of $\mbox{\boldmath $x$}$ and $\mbox{\boldmath $y$}$, respectively. 
\begin{enumerate}[noitemsep]
\item [(a)] The vector $\mbox{\boldmath $x$}$ is said to weakly supermajorize the vector $\mbox{\boldmath $y$}$
 (written as $\mbox{\boldmath $y$}\stackrel{ w}{\preceq} \mbox{\boldmath $x$}$) if
 \begin{eqnarray*}
 \sum\limits_{i=1}^j x_{(i)}\leq \sum\limits_{i=1}^j y_{(i)},\quad \text{for}\;j=1,2,\dots,n;
\end{eqnarray*}
 \item [(b)] The vector $\mbox{\boldmath $x$}$ is said to $p$-larger than the vector $\mbox{\boldmath $y$}$
 (written as $\mbox{\boldmath $y$}\stackrel{ p}{\preceq} \mbox{\boldmath $x$}$) if
 \begin{eqnarray*}
  \prod\limits_{i=1}^j x_{(i)}\leq \prod\limits_{i=1}^j y_{(i)},\quad \text{for}\;j=1,2,\dots,n;
 \end{eqnarray*}
 \item [(c)] The vector $\mbox{\boldmath $x$}$ is said to reciprocally majorize the vector $\mbox{\boldmath $y$}$
 (written as $\mbox{\boldmath $y$}\stackrel{ rm}\preceq \mbox{\boldmath $x$}$) if
 \begin{eqnarray*}
 ~~~~~~~~~~~~~~~~~~~~~~~~~~~~~~ \sum\limits_{i=1}^j \frac{1}{x_{(i)}}\geq \sum\limits_{i=1}^j \frac{1}{y_{(i)}},\quad \text{for}\;j=1,2,\dots,n.~~~~~~~~~~~~~~~~~~~~~~~~~~\hfill\Box
 \end{eqnarray*}
\end{enumerate}
\end{d1}
 \hspace*{0.2 in}Stochastic ageing concepts are very useful tools for describing how a system {ages} over time. In the literature, different ageing classes (such as IFR, DFR, DLR, and so on) have been introduced to characterize different ageing properties of a system (see \cite{BP}). Below, we give the definitions of some ageing classes that are most pertinent to the {ensuing discussions}. 
  \begin{d1}
Let $X$ be an absolutely continuous random variable with nonnegative support. Then, $X$ is said to have
\begin{enumerate}[noitemsep]
\item [$(a)$] increasing likelihood ratio (ILR) (resp. decreasing likelihood ratio (DLR)) property if 
$f'_X(x)/f_X(x)$ is decreasing (resp. increasing) in $x\geq 0;$
\item [$(b)$] increasing failure rate (IFR) (resp. decreasing failure rate (DFR)) property if 
$r_X(x)$ is increasing (resp. decreasing) in $x\geq 0;$
\item [$(c)$] decreasing reversed failure rate (DRFR) property if 
$\tilde r_X(x)\;\text{is decreasing in}\;x\geq 0;$ 
\end{enumerate}
\end{d1}
\section{Ordered random vectors}\label{se111a}
In this section, we give the definition of DSOS and discuss its important special cases. As an extension of the sequential order statistics (SOS), Baratnia and Doostparast~\cite{BD} introduced the developed sequential order statistics (DSOS), which are useful for modelling the lifetime of a system with dependent components. The definition of DSOS is as follows (see ~\cite{BD,NB}).
 \begin{d1}\label{11def1}
Let {$F_1, \dots ,F_{n}$} be $n$ absolutely continuous cumulative distribution functions with $F_1^{-1} (1)\leq \dots \leq  F_n^{-1}(1)$. Consider a system of $n$ components installed at time $t = 0$. Assume that all components of the system are functioning at the {starting time}. Let $X^{(1)}_1, \dots , X^{(1)}_n$ be $n$ dependent and identical (DID) random variables, with distribution functions $F_1$, representing the lifetimes of $n$ components. Assume that the dependence structure between these random variables is described by the Archimedean copula with generator $\phi$. Then, the first component failure time is given by
$$X^{\star}_{1:n} = \min\lt\{X_1^{(1)}, \dots, X_n^{(1)}\rt\}.$$
Given $X^{\star}_{1:n} = t_1$, the residual lifetimes of {the remaining} $(n - 1)$ components are equal in distribution to the residual lifetimes of $(n-1)$ DID components with age $t_1$ and with cumulative distribution {function} $F_2 $, (instead of $F_1$) with the same dependence structure; here, $F_2 $ is assumed in place of $F_1 $ as the failure of the first component {would have} an impact on the performance of other components.
 Let the lifetimes of these DID  components be represented by $X^{(2)}_1, \dots , X^{(2)}_{n-1}$. Then, for $j=1,\dots,n-1$,  $X^{(2)}_j \sim F_2(\cdot | t_1)$, where $\bar{F}_2(x|t_1) = {\bar{F}_2(x)}/{\bar{F}_2(t_1)}$, for $x \geq t_1$. Moreover, $X^{(2)}_j \geq t_1$, for $j = 1, \dots ,n - 1$. {Next}, the second component failure time is given by
$$X^{\star}_{2:n} = \min\lt\{X_1^{(2)}, \dots, X_{n-1}^{(2)}\rt\}.$$
By proceeding in this manner, we assume that the $i$-th failure occurs at time $t_i \;(>t_{i-1})$, i.e., $X^{\star}_{i:n} = t_i$.  Then, the residual lifetimes of {the remaining} $(n - i)$ components are equal in distribution to the residual lifetimes of $(n-i)$ DID components with age $t_i$ and with distribution functions $F_{i+1}$ with the same dependence structure. Let the lifetimes of these DID components be represented by {$X^{(i+1)}_1, \dots , X^{(i+1)}_{n-i}$}. Then, for $j=1,\dots n-i$, $X^{(i+1)}_j \sim F_{i+1}(\cdot | t_i)$, where $\bar{F}_{i+1}(x|t_i) = {\bar{F}_{i+1}(x)}/{\bar{F}_{i+1}(t_i)}$, for $x \geq t_i$. Moreover, note that $X^{(i+1)}_j \geq t_i$, for $j = 1, \dots ,n - i$. Then, the $(i+1)$-th component failure time is given by
$$X^{\star}_{i+1:n} = \min\lt\{X_1^{(i+1)}, \dots, X_{n-i}^{(i+1)}\rt\}.$$
Finally, if the $(n - 1)$-th component failure occurs at time
$t_{n-1} = X^{\star}_{n-1:n}$, then the last component failure time is given by $X^{\star}_{n:n}$ with reliability
function $\bar{F}_n(x|t_{n-1}) = {\bar{F}_n(x)}/{\bar{F}_n(t_{n-1})}$, for $x \geq t_{n-1}$.
Then, the random variables $X^{\star}_{1:n} \leq \dots \leq X^{\star}_{n:n} $ are called developed sequential order statistics (DSOS) based on {$F_1, \dots, F_n$}, where the dependence structure is described by the Archimedean copula with generator $\phi$. In short, we denote them by $(X^{\star}_{1:n}, \dots,X^{\star}_{n:n} )\sim$ DSOS($ F_1  , \dots , F_n;\phi $).  
\end{d1}
\begin{r1}\label{re116}
One may note that, if {$(X^{\star}_{1:n}, \dots,X^{\star}_{n:n} )\sim$ DSOS($ F_1 , \dots , F_n;\phi $)}, then {$\{X^{\star}_{1:n},\dots ,\\X^{\star}_{n:n}\}$} forms a Markov chain with transition {probabilities}
\begin{eqnarray}\label{10def1a1}
P\lt(X^{\star}_{r:n} > t | X^{\star}_{r-1:n} =x\rt) &=& \phi\lt( \lt(n-r+1 \rt) \psi \lt( \frac{\bar{F}_r(t)}{\bar{F}_r(x)} \rt) \rt), \quad t \geq x,\; \bar{F} (x) >0,
\end{eqnarray}
where $\psi\equiv\phi^{-1}$.
$\hfill\Box$
\end{r1}
\hspace*{0.2 in} Generalized order statistics (GOS), a unified notion of ordered random variables, contain many popular models as particular cases, including sequential order statistics (SOS) under PHR model, order statistics with non-integral sample size, $k$-record values, Pfeifer's record values,  $k_n$-records from non-identical distributions, and ordered random variables from truncated distributions. We now give the definition of developed generalized order statistics (DGOS), which is a generalization of GOS (see \cite{CK2, HZ, K}). 
	\begin{d1}\label{l111}
	Let $n \in \mathcal{N}$, $\gamma_{n,n} = \alpha_n = k>0$, $m_1, \dots, m_{n-1} \in \mathcal{R}$, $ M_i = \sum_{j=i}^{n-1} m_j $, $1 \leq i \leq n-1$, $\gamma_{i,n} = k+n-i+M_i  = \lt( n-i +1 \rt) \alpha_i >0$, for all $i = 1, \dots, n-1$, and let $\tilde{m} = (m_1, \dots, m_{n-1})$, $n=2,\dots, n-1$. 
	The random variables {$X\lt ( 1, n, \tilde{m}_{n}, k \rt), \dots, X\lt( n, n, \tilde{m}_{n}, k \rt)$} are said to be developed generalized order statistics (DGOS) from an absolutely continuous distribution function $F$ with probability density function $f$ and dependence structure described by the Archimedean copula with generator $\phi$, denoted by
$\lt(X\lt( 1, n, \tilde{m}_{n}, k \rt), \right.$  $\left. \dots, X\lt( n, n, \tilde{m}_{n}, k \rt)\rt) \sim$ DGOS($ F, \gamma_{1,n}, \dots, \gamma_{n,n}; \phi $),  if their joint probability density function is given by
		\begin{eqnarray*}\label{l111a1}
		f_{ X\lt( 1, n, \tilde{m}_{n}, k \rt), \dots,X\lt( n, n, \tilde{m}_{n}, k \rt) } \lt( x_1,  \dots , x_n \rt) &=& \prod_{j=1}^{n} \lt\{\phi '  \lt( \lt(n-j+1 \rt) \psi \lt(\frac{ \bar{F}^{\alpha_j} \lt( x_j \rt) } { \bar{F}^{\alpha_j} \lt( x_{j-1} \rt) } \rt) \rt) \nonumber
		\right. \\&&\left. 
		\lt(n-j+1 \rt)\alpha_j \psi ' \lt(\frac{ \bar{F}^{\alpha_j} \lt( x_j \rt) } { \bar{F}^{\alpha_j} \lt( x_{j-1} \rt) } \rt) \frac{ \bar{F}^{\alpha_j-1} \lt( x_{j} \rt) {f} \lt( x_j \rt) } { \bar{F}^{\alpha_j} \lt( x_{j-1} \rt) } \rt\}, 
		\end{eqnarray*}
		where $0 =x_0  < \dots < x_n$.     $\hfill\Box$
	\end{d1}
		\hspace*{0.2 in}Like GOS, DGOS also contains many popular models of ordered random variables with dependence structure described by the Archimedean copula, {as listed} in Table~\ref{tb1}.
\begin{center}
\captionof{table}{Models of ordered random variables and their relations with DGOS (see, \cite{CK, CK2}).}\label{tb1}
\begin{tabular}{ |m{3 cm}| m{1.6 cm}| m{1.6 cm}| m{2.5 cm} | m{6.5 cm}| }
\hline
   \multicolumn{2}{|c|}{$\bm{\gamma_{r,n}(1\leq r \leq n-1)}$  } & {\bm{$\gamma_{n,n}}$} & {\bf Dependance structure  ($\phi$)} & \multirow{2}{*}{\bf DGOS Model}   \\ 
   \cline{1-3}
 {$\bm{n-r+1}$}& {$\bm{\alpha_r}$}&  {$\bm{k}$} & &
  \\ \hline
  $n-r+1$ & $1$ & $1$ & $\phi$  & Ordinary order statistics (OS) \\\hline
 $a-r+1: a \in (0,\infty)$ & $1$ & $a-n+1$ & $\phi$  & OS with non integral sample size~\cite{RL} \\\hline
$n-r+1$ & $\alpha_r$ & $\alpha_n$ & $\phi$ & DSOS  under PHR model \\\hline
$n-r+1$ & $\alpha_r$ & $k$ & $\phi(u) = e^{-u}$ & Generalized order statistics (GOS) \\\hline
$1$ & $1$ & $1$ & Not applicable & Record value~\cite{C} \\\hline
 $k$ & $1$ & $k$ & $\phi$ & $k$-th record value~\cite{DK} \\\hline
$1$ & $\alpha_r$ & $\alpha_n$ & Not applicable & Pfeifer's record value~\cite{P}  \\\hline
$k_r$ & $\alpha_r$ & $\alpha_n k_n$ & $\phi$ & Ordering via truncation~\cite{K}  \\\hline
$\nu-r+1$, if $1\leq r \leq r_1$, $\nu-n_1-r+1$, if $r_1 < r \leq n$ & $1$ & $\nu-n_1-n+1$ &  $\phi$ & Progressively type-II censored order statistics~\cite{BC}  \\\hline
\end{tabular}
\end{center}
			\hspace*{0.2 in} Below, we give a list of models containing DSOS and its particular cases in Table~\ref{tbb2}.
			In subsequent sections,  we  discuss various results for these models.
			\newpage
\begin{center}
\captionof{table}{Models of ordered random variables obtained from DSOS}\label{tbb2}
\begin{tabular}{ |m{5 cm}| m{6.53 cm}| m{4.7 cm}| }
\hline
 \text{\bf Condition } & {\bf Notation}& {\bf Model specification} \\ \hline
  NULL & $(X^{\star}_{1:n}, \dots,X^{\star}_{n:n} ) \!  \sim \! $ DSOS($F_1 , F_2 $ $\dots , F_{n} ;\phi$)& DSOS\\\hline
 $\phi(u)=e^{-u}, \;u>0$& $(X^{\star}_{1:n}, \dots,X^{\star}_{n:n} )\sim$ SOS($F_{1} , F_2 $  $ \dots , F_{n}$)& SOS\\
 \hline
 $F_i\sim $PHR($F,\alpha_i)$, for $i=1 ,\dots, n$ &  $(X\lt( 1, n, \tilde{m}_{n}, k \rt), \dots, X\lt( n, n, \tilde{m}_{n}, k \rt))\sim$ DGOS($ F, \gamma_{1,n}, \dots, \gamma_{n,n};\phi $) & GOS with dependent components\\   
  \hline
 $F_i\sim $PHR($F,\alpha_i)$, for $i=1,\dots, n$, and $\phi(u)=e^{-u}, \;u>0$&  $(X\lt( 1, n, \tilde{m}_{n}, k \rt), \dots, X\lt( n, n, \tilde{m}_{n}, k \rt)) \! \sim \; $ GOS($ F, \gamma_{1,n}, \dots, \gamma_{n,n}$) & GOS with independent components\\ 
  \hline
 $F_i= F$, for all $i=1,\dots, n$ &$ (X^{\star}_{1:n}, \dots,X^{\star}_{n:n} )\sim$ OS($ F;\phi $) & OS with DID components\\\hline
\end{tabular}
\end{center}
	\hspace*{0.2 in}We first present some lemmas {that are essential for } proving the main results of this paper.
	\begin{lem}\label{l116}
			Let $(X^{\star}_{1:n}, \dots,X^{\star}_{n:n} )\sim$ DSOS($ F_1 , $ $\dots , F_n;\phi $), {and} $D_i\lt(\cdot\rt)\equiv -\ln\bar{F}_i\lt(\cdot\rt)$ be the cumulative hazard rate function of $F_i$, for $i=1,\dots,n$.
			  Then,
			\begin{eqnarray}\label{l1011}
				&&X_{1:n}^{\star} = D_1^{-1}\lt(W^{(1)}\rt),
				\\&&X_{i:n}^{\star} = D_i^{-1}\lt(W^{(i)} + D_i\lt(X_{i-1:n}^{\star}\rt)\rt), \quad\text{ for } i=2, \dots, n,\label{l1012}
			\end{eqnarray}
		where $$W^{(i)} = -\ln \lt(V^{(i)}\rt)= \min \lt\{-\ln\lt(1-U^{(i)}_1\rt), \dots , -\ln \lt(1-U^{(i)}_{n-i+1}\rt)\rt\}, \quad i=1,\dots, n,$$ 
and $U_j^{i} \sim Unif(0,1)$, for $i=1,\dots, n$, and $j=1, \dots, n-i+1$; here, for each $i\in\{1,\dots, n\}$, $U_j^i$'s are dependent random variables governed by the Archimedean copula with generator $\phi$. Moreover, $\{W^{(i)},\; i=1, \dots , n\}$ are independent with survival {functions}
		\begin{eqnarray}\label{l101a1}
		\bar{F}_{W^{(i)}} \lt(t\rt) 
		&=& \phi\lt( \lt(n-i+1\rt) \psi \lt( e^{-t}\rt) \rt), \quad t>0, i=1,\dots,n,\;\psi\equiv\phi^{-1}.
		\end{eqnarray}
	\end{lem}
\hspace*{0.2 in}The following lemma follows from Remark~\ref{re116} and Lemma~\ref{l116} (see also \cite{CK2, HZ}). 
 \begin{lem}\label{l117}
Let $\lt(X\lt( 1, n, \tilde{m}_{n}, k \rt), \dots, X\lt( n, n, \tilde{m}_{n}, k \rt)\rt) \sim$ DGOS($ F, \gamma_{1,n}, \dots, \gamma_{n,n};\phi $), {and} $D\lt(\cdot\rt)\equiv -\ln\bar{F}\lt(\cdot\rt)$ be the cumulative hazard rate function of $F$. Then, 
 \begin{eqnarray}\label{11def2a1}
\lt( X\lt( 1, n, \tilde{m}_{n}, k \rt), \dots, X\lt( n, n, \tilde{m}_{n}, k \rt) \rt) 
\stackrel{d}= \lt(  D^{-1} \lt( B_{1,n} \rt), \dots, D^{-1} \lt( \sum\limits_{j=1}^nB_{j,n} \rt) \rt) ,
\end{eqnarray}
where 
$$B_{j,n} = \min \lt\{-\frac{1}{\alpha_{j}} \ln\lt(1-U^{(j)}_1\rt), \dots , -\frac{1}{\alpha_{j}} \ln \lt(1-U^{(j)}_{n-j+1}\rt)\rt\} = \frac{1}{\alpha_j} W^{(j)}, \quad j=1,\dots, n,$$
and $U_j^i$'s are as {given} in Lemma~\ref{l116}.
Moreover, the survival function of $B_{j,n}$ {is}
		\begin{eqnarray}\label{11def2a2}
		\bar{F}_{B_{j,n}} \lt(t\rt)  
		&=& \phi\lt( \lt(n-j+1\rt) \psi \lt( e^{-\alpha_{j} t}\rt) \rt), \quad t>0, j=1,\dots,n,\;\psi\equiv\phi^{-1}.
		\end{eqnarray}
\end{lem}
	\section{{Comparing} random vectors from DSOS model with identical components}\label{se113}
	In this section, we establish some stochastic comparison results for {random vectors with} DSOS model in both one-and two-sample {situations}.
	\subsection{One-sample {situation}}\label{sub111}
	In the following three theorems, we compare two random vectors from DSOS model with respect to the usual multivariate stochastic and dynamic multivariate hazard rate orders. We provide the results in light of the assumptions made on the underlying distribution functions upon which the DSOS models are {built}. We only give the proof of Theorem~\ref{t116} and the proofs of Theorems~\ref{t114} and \ref{t115} are omitted for the sake of brevity.
	\begin{thm}\label{t114}
 Let $\lt(X^{\star}_{1:n}, \dots,X^{\star}_{n:n} \rt)\sim$ DSOS($ F_1 , \dots , F_n;\phi $) and $(X^{\star}_{1:n+1}, \dots, X^{\star}_{n+1:n+1} )\sim$ $  $ DSOS($ F_1 , \dots , F_{n+1};\phi $). Then, the following results hold true{:}
	\begin{enumerate}
	\item [$(a)$] $\lt( X^{\star}_{1:n+1}, \dots,X^{\star}_{n:n+1} \rt)  \; \leq_{st} \;  \lt( X^{\star}_{1:n}, \dots,X^{\star}_{n:n} \rt)$; 
	\item [$(b)$] {Suppose} $uR'(u)/R(u)$ is positive and increasing in $u>0$. Then{,} $( X^{\star}_{1:n+1},  \dots, X^{\star}_{n:n+1} ) $  $\leq_{dyn-hr} \;  \lt( X^{\star}_{1:n},  \dots,X^{\star}_{n:n} \rt)$. 
	\end{enumerate}	
	\end{thm}
	\begin{thm}\label{t115}
Let $(X^{\star}_{1:n}, \dots,X^{\star}_{n:n} )\sim$ DSOS($ F_1 , \dots , F_n;\phi $).
Then, the following results hold true{:}
	\begin{enumerate}
	\item [$(a)$] If $F_2  \leq_{hr} \dots \leq_{hr} F_n$, then $\lt( X^{\star}_{1:n}, \dots,X^{\star}_{n-1:n} \rt)  \; \leq_{st} \;  \lt( X^{\star}_{2:n}, X^{\star}_{3:n}, \dots,X^{\star}_{n:n} \rt)$; 
	\item [$(b)$] {Suppose} $uR'(u)/R(u)$ is positive and increasing in $u>0$. If $F_1 \leq_{hr} \dots \leq_{hr} F_n$, then 
$\lt( X^{\star}_{1:n},  \dots,X^{\star}_{n-1:n} \rt)  \; \leq_{dyn-hr} \;  \lt( X^{\star}_{2:n}, \dots,X^{\star}_{n:n} \rt)$. 
	\end{enumerate}
	\end{thm}
	\begin{thm}\label{t116}
Let $(X^{\star}_{1:n}, \dots,X^{\star}_{n:n} )\sim$ DSOS($ F_1 ,  \dots , F_n;\phi $) and $(X^{\star}_{1:n},  \dots, X^{\star}_{n+1:n+1} )$ $\sim$ DSOS($ F_1 , $  $ \dots , F_{n+1};\phi $).
Then, the following results hold true{:}
	\begin{enumerate}
	\item [$(a)$] If $F_1 \leq_{st} F_2 \leq_{hr} \dots \leq_{hr} F_{n+1}$, then $\lt( X^{\star}_{1:n}, \dots,X^{\star}_{n:n} \rt)  \; \leq_{st} \;  \lt( X^{\star}_{2:n+1}, \dots,X^{\star}_{n+1:n+1} \rt)$; 
	\item [$(b)$] {Suppose} $uR'(u)/R(u)$ is increasing in $u>0$. If $F_1 \leq_{hr}  \dots \leq_{hr} F_{n+1}$, then 
$\lt( X^{\star}_{1:n}, \dots,X^{\star}_{n:n} \rt)  $  $ \leq_{dyn-hr} \;  \lt( X^{\star}_{2:n+1}, \dots,X^{\star}_{n+1:n+1} \rt)$. 
	\end{enumerate}
	\end{thm}	
\subsection{Two-sample {situation}}\label{sub112}
	In the following theorem, we compare two random vectors with DSOS model {that} are formed from two different samples, with respect to the usual multivariate stochastic, dynamic multivariate hazard rate and multivariate dispersive orders. 
	{The proof of the second part follows {along} the same {lines as those of} the first part and {is} therefore omitted}.
	\begin{thm}\label{t117}
Let $\lt(X^{\star}_{1:n}, \dots,X^{\star}_{n:n} \rt)\sim$ DSOS($ F_1 , \dots , F_n;\phi $) and $\lt(Z^{\star}_{1:n}, \dots,Z^{\star}_{n:n} \rt)\sim$ DSOS($ G_1 , $   $ \dots , G_{n};\phi $).
 Then, the following results hold true{:}
 \begin{enumerate}
         \item [$(a)$]  {Suppose} $uR'(u)/R(u)$ is increasing in $u>0$. If $F_1 \leq_{st} G_1$ and $F_i \leq_{hr} G_i$, $i= 2, \dots, n$, then $\lt( X^{\star}_{1:n}, \dots,X^{\star}_{n:n} \rt)  \; \leq_{st} \;  \lt( Z^{\star}_{1:n}, \dots,Z^{\star}_{n:n} \rt)$; 
	\item [$(b)$] {Suppose} $uR'(u)/R(u)$ is increasing in $u>0$. If $F_i \leq_{hr} G_i$, $i=1, \dots, n$, then $\lt( X^{\star}_{1:n}, \dots,X^{\star}_{n:n} \rt) $ $ \; \leq_{dyn-hr} \;  \lt( Z^{\star}_{1:n}, \dots,Z^{\star}_{n:n} \rt)$; 
	\item [$(c)$] Let $F_1\stackrel{d}=\dots\stackrel{d}=F_n\stackrel{d}=F$ and $G_1\stackrel{d}=\dots\stackrel{d}=G_n\stackrel{d}=G$. If $F \leq_{disp} G$, then 
$\lt( X^{\star}_{1:n},  \dots,X^{\star}_{n:n} \rt)  \; \leq_{disp} \;  \lt( Z^{\star}_{1:n}, \dots,Z^{\star}_{n:n} \rt)$. 
	\end{enumerate}
	\end{thm}
	\section{{ Comparing random vectors from} DGOS model with identical components}\label{se114}
		\hspace*{0.2 in}In this section, we establish some stochastic comparisons {of random vectors with DGOS model when the underlying components are identical}. 
\subsection{One-sample {situation}}\label{sub113a}
	In the following theorem, we compare random vectors with DGOS model with respect to the multivariate dispersive order.
	\begin{thm}\label{t1114}
	Let $(X\lt( 1, n, \tilde{m}_{n}, k \rt),  \dots, X\lt( n, n, \tilde{m}_{n}, k \rt))\sim$ 
	DGOS($ F, \gamma_{1,n},  \dots,$ $ \gamma_{n,n};\phi $) {be} such that $F$ is DFR, $ \tilde{m}_{n+1} = \lt(  \tilde{m}_{n}, m_n\rt)$, for $n \in \mathcal{N}$, and $m_i +1\geq 0$ for each $i$. 
	Then, the following results hold true{:}
	\begin{enumerate}
	\item [$(a)$] {Suppose} $R(u) $ is decreasing in $u>0$. If $m_n \leq \min\lt\{ m_1, \dots, m_{n-1}\rt\}$,
	then $\lt( 0, X\lt( 1, n, \tilde{m}_{n}, k \rt), \right.$  $\left. \dots,X\lt( n-1, n, \tilde{m}_{n}, k \rt) \rt)  \; \leq_{disp} 
	  ( X( 1, n, \tilde{m}_{n}, k ), \dots,$ $X( n, n, \tilde{m}_{n}, k )) $;
	\item [$(b)$] {Suppose} $R(u) $ is decreasing in $u>0$. Then, $(  X\lt( 1, n+1, \tilde{m}_{n+1}, k \rt),\dots,$ $X\lt( n, n+1, \tilde{m}_{n+1}, k \rt) )  
	\\ \leq_{disp} \; 
	\lt( X\lt( 1, n, \tilde{m}_{n}, k \rt), \dots, X\lt( n, n, \tilde{m}_{n}, k \rt) \rt)$;
	\item [$(c)$] If $m_n \leq \min\lt\{ m_1, \dots, m_{n-1}\rt\}$, then 
	{$\lt( 0, X\lt( 1, n, \tilde{m}_{n}, k \rt), \dots,X\lt( n, n, \tilde{m}_{n}, k \rt) \rt)  
	 \leq_{disp} 
	\\ \lt( X\lt( 1, n+1, \tilde{m}_{n}, k \rt),  \dots, X\lt( n+1, n+1, \tilde{m}_{n+1}, k \rt) \rt)$}.  $\hfill\Box$
	\end{enumerate}
	\end{thm}
	\hspace*{0.2 in}In the following theorem, we {discuss} multivariate dispersive ordering for multivariate marginals from the DGOS model. 
	\begin{thm}\label{t1115}
	Let $(X\lt( 1, n, \tilde{m}_{n}, k \rt),  \dots, X\lt( n, n, \tilde{m}_{n}, k \rt))\sim$ 
	DGOS($ F, \gamma_{1,n},$ $ \dots, \gamma_{n,n};\phi $) {be} such that $F$ is DFR, $ \tilde{m}_{n+1} = \lt(  \tilde{m}_{n}, m_n\rt)$, for $n \in \mathcal{N}$, and $m_i +1\geq 0$ for each $i$. Further, let $1 \leq p_1 < \dots < p_i \leq n$, and $1\leq q_1 < \dots < q_i \leq n$ {be} such that $q_i - p_i  \geq \dots \geq q_1 - p_1 \geq 0$. {Suppose} ${G(nu)}/{R(u)} - {G(u)}/{R(u)} $ is positive and increasing in $u>0$.
	Then, the following results hold true{:}
	\begin{enumerate}
	\item [$(a)$] If ${R(u)} $ is decreasing in $u>0$ and $m_n \leq \min\lt\{ m_1, \dots, m_{n-1}\rt\}$,
	then
	$( X( p_1, n, \tilde{m}_{n}, k ), \dots,$  $X( p_i, n, \tilde{m}_{n}, k ) )  \; \leq_{disp} \;  ( X( q_1, n, \tilde{m}_{n}, k ), \dots, X( q_i, n, \tilde{m}_{n}, k ) )$;
	\item [$(b)$] If  $R(u) $ is decreasing in $u>0$ and $m_n \leq \min\lt\{ m_1, \dots, m_{n-1}\rt\}$,
	then 
	$\lt(  X\lt( p_1, n+1, \tilde{m}_{n+1}, k \rt),\right.$   $\left. \dots,X\lt( p_i, n+1, \tilde{m}_{n+1}, k \rt) \rt)  \; \leq_{disp} \;  \lt( X\lt( q_1, n, \tilde{m}_{n}, k \rt), \dots, X\lt( q_i, n, \tilde{m}_{n}, k \rt) \rt)$;
	\item [$(c)$] If $m_n \leq \min\lt\{ m_1, \dots, m_{n-1}\rt\}$, then 
	$\lt( X\lt( p_1, n, \tilde{m}_{n}, k \rt), \dots,X\lt( p_i, n, \tilde{m}_{n}, k \rt) \rt)  \; \leq_{disp} \;  
	\\\lt( X\lt( q_1, n+1, \tilde{m}_{n+1}, k \rt), \dots, X\lt( q_i, n+1, \tilde{m}_{n+1}, k \rt) \rt)$.  $\hfill\Box$
	\end{enumerate}
	\end{thm}
	\hspace*{0.2 in}In the following theorems, we present some univariate results. We compare two DGOS models with respect to {the usual stochastic, hazard rate, reverse hazard rate, likelihood ratio and dispersive orders}. 
	Theorem~\ref{t1111a}(a) is trivially true, {while parts (b) and (c) of Theorem~\ref{t1111a}} follow from Theorem~\ref{t114}(a) and Theorem~\ref{t116}(a), respectively. 
	For the sake of brevity, the proofs of Theorems~\ref{t1111}(a) and (c),~\ref{t1112}(a) and (b),  \ref{t1110}(a) and (c), {and} \ref{t1113} (a) and (c) are omitted. 
		\begin{thm}\label{t1111a}
	Let $(X\lt( 1, n, \tilde{m}_{n}, k \rt), \dots, X\lt( n, n, \tilde{m}_{n}, k \rt))\sim$ 
	DGOS($ F, \gamma_{1,n}, \dots,$ $ \gamma_{n,n};\phi $), $ \tilde{m}_{n+1} = \lt(  \tilde{m}_{n}, m_n\rt)$, for $n \in \mathcal{N}$, and $m_i +1\geq 0$ for each $i$. Then, the following results hold true{:}
	\begin{enumerate}
	\item [$(a)$]  $X\lt( i, n, \tilde{m}_{n}, k \rt) \; \leq_{st} \; X\lt( i+1, n, \tilde{m}_{n}, k \rt) $, for $i= 1, \dots, n-1$;
	\item [$(b)$]   $X\lt( i, n+1, \tilde{m}_{n+1}, k \rt) \; \leq_{st} \; X\lt( i, n, \tilde{m}_{n}, k \rt)$, for $i= 1,\dots, n$;
	\item [$(c)$]  If $m_n \leq \min\lt\{ m_1, \dots, m_{n-1}\rt\}$, then $X\lt( i, n, \tilde{m}_{n}, k \rt) \; \leq_{st} \; X\lt( i+1, n+1, \tilde{m}_{n+1}, k \rt)$, for $i= 1, \dots, n$.  
	\end{enumerate}
	\end{thm}
	\begin{thm}\label{t1111}
	Let $(X\lt( 1, n, \tilde{m}_{n}, k \rt),  \dots, X\lt( n, n, \tilde{m}_{n}, k \rt))\sim$ 
	DGOS($ F, \gamma_{1,n}, \dots, $ $\gamma_{n,n};\phi $), $ \tilde{m}_{n+1} = \lt(  \tilde{m}_{n}, m_n\rt)$, for $n \in \mathcal{N}$, and $m_i +1\geq 0$ for each $i$. Then, the following results hold true{:}
	\begin{enumerate}
	\item [$(a)$] {Suppose} $uR'(u)/R(u)$ is increasing in $u>0$. Then{,} $X\lt( i, n, \tilde{m}_{n}, k \rt) \; \leq_{hr} \; X\lt( i+1, n, \tilde{m}_{n}, k \rt) $, for $i= 1, \dots, n-1$;
	\item [$(b)$] {Suppose} $uR'(u)/R(u)$ is increasing and positive in $u>0$. Then{,} $X\lt( i, n+1, \tilde{m}_{n+1}, k \rt) \; \leq_{hr} \; X\lt( i, n, \tilde{m}_{n}, k \rt)$, for $i= 1,\dots, n$;
	\item [$(c)$] {Suppose} $uR'(u)/R(u)$ is increasing in $u>0$. If $m_n \leq \min\lt\{ m_1, \dots, m_{n-1}\rt\}$, then {$X\lt( i, n, \tilde{m}_{n}, k \rt) \; 
	\\ \leq_{hr} \; X\lt( i+1, n+1, \tilde{m}_{n+1}, k \rt)$, for $i= 1, \dots, n$}.  
	\end{enumerate}
	\end{thm}
	\begin{thm}\label{t1112}
	Let $(X\lt( 1, n, \tilde{m}_{n}, k \rt), \dots, X\lt( n, n, \tilde{m}_{n}, k \rt))\sim$ 
	DGOS($ F, \gamma_{1,n},  \dots, $ $\gamma_{n,n};\phi $), $ \tilde{m}_{n+1} = \lt(  \tilde{m}_{n}, m_n\rt)$, for $n \in \mathcal{N}$, and $m_i +1\geq 0$ for each $i$. Then, the following results hold true{:}
	\begin{enumerate}
	\item [$(a)$] {Suppose} $uH'(u)/H(u)$ is decreasing in $u>0$. Then{,} $X\lt( i, n, \tilde{m}_{n}, k \rt) \; \leq_{rh} \; X\lt( i+1, n, \tilde{m}_{n}, k \rt)$, for $i= 1, \dots, n-1$;
	\item [$(b)$] {Suppose} $uH'(u)/H(u)$ is decreasing and negative in $u>0$. Then{,} $X\lt( i, n+1, \tilde{m}_{n+1}, k \rt) 
	$  $\; \leq_{rh} \; X\lt( i, n, \tilde{m}_{n}, k \rt)$, for $i= 1, \dots, n$;
	\item [$(c)$] {Suppose} $uH'(u)/H(u)$ is decreasing in $u>0$. If $m_n \leq \min\lt\{ m_1, \dots, m_{n-1}\rt\}$, then {$X\lt( i, n, \tilde{m}_{n}, k \rt) \; 
	\\ \leq_{rh} \; X\lt( i+1, n+1, \tilde{m}_{n+1}, k \rt)$, for $i= 1, \dots, n$}.  
	\end{enumerate}
	\end{thm}
	\begin{thm}\label{t1110}
	Let $(X\lt( 1, n, \tilde{m}_{n}, k \rt), \dots, X\lt( n, n, \tilde{m}_{n}, k \rt))\sim$ 
	DGOS($ F, \gamma_{1,n},  \dots, $ $\gamma_{n,n};\phi $), $ \tilde{m}_{n+1} = \lt(  \tilde{m}_{n}, m_n\rt)$, for $n \in \mathcal{N}$, and $m_i +1\geq 0$ for each $i$. {Suppose} ${G(nu)}/{R(u)} - {G(u)}/{R(u)} $ is positive and increasing in $u>0$.Then, the following results hold true{:}
	\begin{enumerate}
	\item [$(a)$] $X\lt( i, n, \tilde{m}_{n}, k \rt) \; \leq_{lr} \; X\lt( i+1, n, \tilde{m}_{n}, k \rt)$, for $i= 1, \dots, n-1$;
	\item [$(b)$] $X\lt( i, n+1, \tilde{m}_{n+1}, k \rt) \; \leq_{lr} \; X\lt( i, n, \tilde{m}_{n}, k \rt)$, for $i= 1, \dots, n$;
	\item [$(c)$] If $m_n \leq \min\lt\{ m_1, \dots, m_{n-1}\rt\}$, then $X\lt( i, n, \tilde{m}_{n}, k \rt) \; \leq_{lr} \; X\lt( i+1, n+1, \tilde{m}_{n+1}, k \rt)$, for $i= 1, \dots, n$.  
	\end{enumerate}
	\end{thm}
	\begin{thm}\label{t1113}
	Let $(X\lt( 1, n, \tilde{m}_{n}, k \rt), \dots, X\lt( n, n, \tilde{m}_{n}, k \rt))\sim$ 
	DGOS($ F, \gamma_{1,n}, \dots, $ $\gamma_{n,n};\phi $), $ \tilde{m}_{n+1} = \lt(  \tilde{m}_{n}, m_n\rt)$, for $n \in \mathcal{N}$, and $m_i +1\geq 0$ for each $i$. {Suppose} ${G(nu)}/{R(u)} - {G(u)}/{R(u)} $ is positive and increasing in $u>0$.Then, the following results hold true{:}
	\begin{enumerate}
	\item [$(a)$] $X\lt( i, n, \tilde{m}_{n}, k \rt) \; \leq_{disp} \; X\lt( i+1, n, \tilde{m}_{n}, k \rt)$, for $i= 1, \dots, n-1$;
	\item [$(b)$] If $R(u)$ is decreasing in $u>0$, then $X\lt( i, n+1, \tilde{m}_{n+1}, k \rt) \; \leq_{disp} \; X\lt( i, n, \tilde{m}_{n}, k \rt)$, for $i= 1, \dots, n$;
	\item [$(c)$] If $m_n \leq \min\lt\{ m_1, \dots, m_{n-1}\rt\}$, then $X\lt( i, n, \tilde{m}_{n}, k \rt) \; \leq_{disp} \; X\lt( i+1, n+1, \tilde{m}_{n+1}, k \rt)$, for $i= 1, \dots, n$.  
	\end{enumerate}
	\end{thm}
	\subsection{Two-sample {situation}}\label{sub113b}
	We first need the following lemma for proving the main results here.
	\begin{lem}\label{l1112}
	Let $(X\lt( 1, n, \tilde{m}_{n}, k \rt), \dots, X\lt( n, n, \tilde{m}_{n}, k \rt))\sim$ 
	DGOS($ F, \gamma_{1,n}, \dots, $ $\gamma_{n,n};\phi $) and 
	\\$(Y\lt( 1, n, \tilde{m}_{n}, k \rt),  \dots, Y\lt( n, n, \tilde{m}_{n}, k \rt))\sim$ 
	DGOS($ G, \gamma_{1,n}, \dots, \gamma_{n,n};\phi $), where $\gamma_i = (n-i+1) \alpha_i$. Similarly, let $(X\lt( 1, n', \tilde{m}'_{n'}, k \rt),  \dots, X\lt( n', n', \tilde{m}'_{n'}, k \rt)) \sim$ DGOS($ F, \gamma_{1,n}',  \dots, \gamma_{n,n}';\phi $) and 
	\\$(Y\lt( 1, n', \tilde{m}'_{n'}, k \rt), \dots, Y\lt( n', n', \tilde{m}'_{n'}, k \rt)) \sim$ DGOS($ G, \gamma_{1,n}', \dots, \gamma_{n,n}';\phi $), where $\gamma_i' = (n-i+1) \beta_i$. {Suppose} ${G(nu)}/{R(u)} - {G(u)}/{R(u)} $ is positive and increasing in $u>0$. If $\gamma_1' \leq \gamma_1$ and $X\lt( 1, n, \tilde{m}_{n}, k \rt) \leq_{icx} Y\lt( 1, n, \tilde{m}_{n}, k \rt)$, then $X\lt( 1, n', \tilde{m}'_{n'}, k \rt) \leq_{icx} Y\lt( 1, n', \tilde{m}'_{n'}, k \rt)$. $\hfill\Box$
	\end{lem}
		\hspace*{0.2 in} In the following theorem, two DGOS models are compared with respect to increasing convex order.  {By using the above lemma, the proof of this theorem can be presented along the same {lines as those} in Theorem 3.11 of Balakrishnan et al.~\cite{BBSS} and is therefore omitted.}
	\begin{thm}\label{t1117}
	Let $(X\lt( 1, n, \tilde{m}_{n}, k \rt), \dots, X\lt( n, n, \tilde{m}_{n}, k \rt))\sim$ 
	DGOS($ F, \gamma_{1,n}, \dots, $ $\gamma_{n,n};\phi $) and 
	\\$(Y\lt( 1, n, \tilde{m}_{n}, k \rt),  \dots, Y\lt( n, n, \tilde{m}_{n}, k \rt))\sim$ DGOS($ G, \gamma_{1,n},  \dots, $ $\gamma_{n,n};\phi $) with $m_i +1\geq 0$, for each $i$. {Suppose} ${G(nu)}/{R(u)} - {G(u)}/{R(u)} $ is positive and increasing in $u>0$. If $X\lt( 1, n, \tilde{m}_{n}, k \rt) \leq_{icx} Y\lt( 1, n, \tilde{m}_{n}, k \rt)$, then $X\lt( i, n, \tilde{m}_{n}, k \rt) \leq_{icx} Y\lt( i, n, \tilde{m}_{n}, k \rt)$, $i=2, \dots, n$. $\hfill\Box$
	\end{thm}
		\hspace*{0.2 in} In the following theorem, we compare two GOS model with respect to {hazard rate, likelihood ratio, dispersive, mean residual life and increasing convex orders}. We only give the proof of part (e) {while proofs of other parts} can be done in the same {way}.
	\begin{thm}\label{t1118}
	Let $(X\lt( 1, n, \tilde{m}_{n}, k \rt), \dots, X\lt( n, n, \tilde{m}_{n}, k \rt)) \! \sim \; $ GOS($ F, \gamma_{1,n},  \dots, \gamma_{n,n}$) and 
	\\$(X\lt( 1, n', \tilde{m}'_{n'}, k' \rt), \dots, X\lt( n', n', \tilde{m}'_{n'}, k' \rt)) \sim$ GOS($ F, \gamma_{1,n}',  \dots, \gamma_{n,n}'$). Let $i\in\{1,2,\dots,n\}$.
		Then, the following results hold true{:}
	\begin{enumerate}[noitemsep]
	\item [$(a)$] If $\lt(\gamma_1, \dots , \gamma_i \rt) \overset{p}{\preceq}  \lt(\gamma_1', \dots , \gamma_i' \rt)$, then $X\lt( i, n, \tilde{m}_{n}, k \rt) \leq_{hr} X\lt( i, n, \tilde{m}'_{n'}, k' \rt)$;
	\item [$(b)$] If $\lt(\gamma_1,  \dots , \gamma_i \rt) \overset{w}{\preceq} \lt(\gamma_1', \dots , \gamma_i' \rt)$, then $X\lt( i, n, \tilde{m}_{n}, k \rt)  \leq_{lr} X\lt( i, n, \tilde{m}'_{n'}, k' \rt)$;
	\item [$(c)$] If $F$ is DFR and $\lt(\gamma_1,  \dots , \gamma_i \rt) \overset{p}{\preceq}  \lt(\gamma_1',  \dots , \gamma_i' \rt)$, then 
	$X\lt( i, n, \tilde{m}_{n}, k \rt) \leq_{disp} X\lt( i, n, \tilde{m}'_{n'}, k' \rt)$;
	\item [$(d)$]  If $F$ is DFR and $\lt(\gamma_1,  \dots , \gamma_i \rt) \overset{rm}{\preceq}  \lt(\gamma_1',  \dots , \gamma_i' \rt)$, then 
	$X\lt( i, n, \tilde{m}_{n}, k \rt) \leq_{mrl} X\lt( i, n, \tilde{m}'_{n'}, k' \rt)$;
	\item [$(e)$]  If $F$ is DFR and $\lt(\gamma_1,  \dots , \gamma_i \rt) \overset{rm}{\preceq}  \lt(\gamma_1',  \dots , \gamma_i' \rt)$, then 
	$X\lt( i, n, \tilde{m}_{n}, k \rt) \leq_{icx} X\lt( i, n, \tilde{m}'_{n'}, k' \rt)$. 
	\end{enumerate}
	\end{thm}
\section{Examples}\label{s117}
%
In this section, we discuss some examples to demonstrate the sufficient conditions given in theorems of the previous sections. Note that these sufficient conditions are satisfied by many popular Archimedean copulas (with specific choices of parameters) that capture both positive and negative dependence structures. 
For the sake of completeness, below we give three examples. Some more examples can be found in \cite{SH, SH1}.
\\\hspace*{0.2 in}The following example demonstrates the condition `` $uR'(u)/R(u)$ is positive and increasing in $u>0$'' 
\begin{ex}\label{ex111}
	Consider the Archimedean copula with generator
\begin{eqnarray*}
	\phi(u)&=&e^{\frac{1}{\theta_1} \lt( 1-e^{u}\rt)},\quad\theta_1 \in\left(0,1\right], \; u>0,
\end{eqnarray*}
which gives
	\begin{eqnarray*}
		\frac{uR'(u)}{R(u)}&=&1+u, \quad u>0.
	\end{eqnarray*}
	It can be easily shown that $uR'(u)/R(u)$ is positive and
increasing in $u > 0$. Thus, the required condition is satisfied.$\hfill\Box$
\end{ex}
\hspace*{0.2 in}Below we give an example that illustrates the  condition `` $uH'(u)/H(u)$ is negative and decreasing in $u>0$''.
\begin{ex}
	Consider the Archimedean copula with generator
	\begin{eqnarray*}
		\phi(u)&=&1-\lt(1-e^{-u}\rt)^{\frac{1}{\theta_2} },\quad \theta_{2} \in\left[1,\infty\right) \; u>0,
		\end{eqnarray*}
which gives
	\begin{eqnarray*}
		\frac{uH'(u)}{H(u)}&=&-\frac{1 + e^u ( u-1)}{ e^u-1}, \quad u>0.
	\end{eqnarray*}
	It can be easily shown that $uH'(u)/H(u)$ is negative and
decreasing in $u > 0$. Thus, the required condition is satisfied.$\hfill\Box$
\end{ex} 
\hspace*{0.2 in}The following example illustrates the condition ``${G(nu)}/{R(u)} - {G(u)}/{R(u)} $ is positive and increasing in $u>0$''.
\begin{ex}
	Consider the Archimedean copula with generator
	\begin{eqnarray*}
		\phi(u)&=&e^{1-\lt(1+u\rt)^{\frac{1}{\theta_{3}}}},\quad \theta_{3} \in\left(0,\infty\right) \; u>0,
	\end{eqnarray*}
	which gives
	\begin{eqnarray*}
		&&G(u)=-\frac{1}{\theta_{3}}u\lt(1+u\rt)^{\frac{1}{\theta_{3}}-1}+u\lt(1+u\rt)^{-1}\lt(\frac{1}{\theta_{3}}-1\rt),\quad u>0,
		\\&&R(u)=-\frac{1}{\theta_{3}}u\lt(1+u\rt)^{\frac{1}{\theta_{3}}-1},\quad u>0,
	\end{eqnarray*}
	and
	\begin{eqnarray*}
		\frac{G(u)}{R(u)}&=&1-\frac{1-\theta_{3}}{\lt(1+u\rt)^{\frac{1}{\theta_{3}}}}, \quad u>0. 
	\end{eqnarray*}	
	Let us fix $\theta_3=0.4, 0.5$ and $0.6$. 
	It can be easily shown that $uG'(u)/G(u)$ and $G(u)/R(u)$ are increasing in $u > 0$. Consequently, from Remark 3.1(a) of Sahoo and Hazra~\cite{SH}, we have that ${G(nu)}/{R(u)} - {G(u)}/{R(u)} $ is positive and increasing in $u>0$.
\end{ex} 
	\section{Concluding remarks}\label{se116}
There are several models of ordered random variables/vectors that arise naturally in practice, such as ordinary order statistics, order statistics with non-integral sample size, $k$-record values, Pfeifer's records,  $k_n$-records from non-identical distributions, ordered random variables from truncated distributions, progressively type-II censored order statistics, and so on. The generalized order statistics (GOS) and sequential order statistics (SOS) are two generalized models that contain all the aforementioned models as particular cases. However, these two models are defined based on the assumption that the underlying random variables are independant. As a generalization, we consider {here} the developed generalized order statistics (DGOS) and developed sequential order statistics (DSOS) models which capture the dependence structure between the underlying random variables. We then establish different univariate and multivariate ordering properties of DSOS and DGOS, {wherein} the dependence structures {between the underlying random variables} are described by the Archimedean copula. 
The results established here generalize many {results} for the GOS and SOS models with identical components {known in the literature}.
	\\\hspace*{0.2 in}The main focus of this paper is to consider the models of ordered random vectors, where the dependence structure is described by the Archimedean copula. The family of Archimedean copulas is popular due to its {flexibility} and ability to describe a wide range of {dependence}. Hence, the study of DSOS and DGOS models governed by the Archimedean copula for comparing the involved ordered random variables is naturally of great interest. 
\\\\{\bf Acknowledgments} 
\vspace*{0.1 in}
 \\\hspace*{0.2 in}The first author sincerely
acknowledges the financial support received from UGC, Govt. of India, {while the} work of the second author was supported by IIT
Jodhpur, India.
\end{document}